\documentclass[usenatbib,usegraphicx,onecolumn]{mn2e}

\def\aj{AJ}
\def\apj{ApJ}
\def\mnras{MNRAS}
\def\aap{A\&A}
\def\apjs{ApJS}
\def\apjl{ApJ Letters}
\def\physrep {Physics Reports}
\begin{document}
\title[Multifractal analysis of SDSS]{The scale of homogeneity of  the galaxy distribution in SDSS DR6 }

\author[Sarkar, Yadav, Pandey \& Bharadwaj]
       {Prakash Sarkar$^1$\thanks{E-mail: prakash@cts.iitkgp.ernet.in},
         Jaswant Yadav$^2$\thanks{E-mail: jaswant@hri.res.in},
         Biswajit Pandey$^3$\thanks{E-mail: biswa@iucaa.ernet.in}
         and
         Somnath Bharadwaj$^1$\thanks{E-mail: somnathb@iitkgp.ac.in}\\
         $^1$Department of Physics \& Meteorology and Centre for Theoretical Studies, Indian Institute of Technology, Kharagpur 721302, India\\
         $^2$Harish Chandra Research Institute, Chhatnag Road, Jhunsi, Allahabad 211019, India\\
         $^3$Inter-University Centre for Astronomy and Astrophysics, Pune 411007, India\\
       }

       \maketitle
       
       \date{\today}
       
       \begin{abstract}
         The assumption that the Universe,  on sufficiently large scales, 
         is homogeneous and isotropic is crucial to our current understanding of 
         cosmology. In this paper  we test if  the observed  galaxy 
         distribution is actually homogeneous on large scales. We have carried out
         a multifractal analysis of the galaxy distribution in a 
         volume limited  subsample from the SDSS DR6. This  considers
         the scaling properties of different moments of  galaxy number counts in
         spheres of varying radius $r$ centered on galaxies. This analysis gives the 
         spectrum of generalized  dimension $D_q(r)$, where $q >0$ quantifies the 
         scaling properties in overdense regions and $q<0$ in underdense regions. 
         We expect  $D_q(r)=3$ for a  homogeneous, random  point distribution.  
         
         In our analysis we have determined $D_q(r)$ in the range $-4 \le q \le
         4$ and $7 \le r \le 98 \,h^{-1} {\rm Mpc}$. In addition to the SDSS
         data we have analysed several random samples which are homogeneous by
         construction. Simulated galaxy samples generated from dark matter
         N-body simulations and the Millennium Run were also analysed. The SDSS
         data is considered to be homogeneous if the measured $D_q$ is
         consistent with that of the random samples. We find that the galaxy
         distribution becomes homogeneous at a length-scale between $60$ and
         $70 \,h^{-1} {\rm Mpc}$.  The galaxy distribution, we find, is
         homogeneous at length-scales greater than $70 \,h^{-1} {\rm Mpc}$.
         This is consistent with earlier works which find the transition to
         homogeneity at around $70 \,h^{-1} {\rm Mpc}$.
       \end{abstract}
       
       
       \section{Introduction}
       The assumption that the universe is homogeneous and isotropic, known
       as the Cosmological Principle, is perhaps the most fundamental
       postulate of the currently well accepted standard model of cosmology
       (e.g. \citealt{weinb}). The Sloan Digital Sky Survey (SDSS), the
       largest galaxy survey till date, has mapped out the galaxy
       distribution in a large volume of space. This provides a unique
       opportunity to observationally test the Cosmological Principle.  The
       two-point correlation function and the power spectrum, which are the
       standard techniques to quantify galaxy clustering
       \citep{1980lssu.book.....P}, rely on the assumption that the
       Cosmological Principle is valid on a sufficiently large scale, and
       hence cannot be used to test the Cosmological Principle.

       \citet{colem} have proposed that the universe has a fractal
       structure. If true, this violates the Cosmological Principle.  While
       some of the subsequent analysis of galaxy surveys show evidence for
       homogeneity in the galaxy distribution at sufficiently large scales
       \citep{1997NewA....2..517G,1999A&A...351..405B,2000MNRAS.318L..51P,
         2000BSAO...50...39T,2001A&A...370..358K}, there are others who either
       fail to find a transition to homogeneity or find definite evidence for
       a fractal structure out to the largest scale probed
       \citep{1999MNRAS.310.1128H, 1999ApJ...514L...1A,2000ApJ...541..519B,
         2004AstL...30..444B}.
       
       In a recent study \citet{2005MNRAS.364..601Y} have analysed nearly
       two-dimensional (2D) strips from the SDSS Data Release One to find a
       definite transition to homogeneity occurring at the length-scale $60 -
       70 \,h^{-1} {\rm Mpc}$.  \citet{2005ApJ...624...54H} have analysed
       the distribution of Luminous Red Galaxies (LRG) in the SDSS to find a
       transition to homogeneity at $\sim 70 \,h^{-1}Mpc$.  The transition to
       homogeneity, however, is contested by
       \citet{2007A&A...465...23S,2009EL.....8529002S,2009EL.....8649001S}
       who fail to find a transition to homogeneity at the largest scales
       $(\sim 100 \,h^{-1} {\rm Mpc})$ which they probe in the 2dF and SDSS
       galaxy surveys.

       In this paper we have analysed the SDSS Data Release Six (DR6) to test
       if the galaxy distribution exhibits a transition to homogeneity at
       large scales, and if so to determine the scale of homogeneity. To
       achieve this we have carried out a multifractal analysis
       \citep{1990MNRAS.242..517M, 1995PhR...251....1B} of the galaxy
       distribution in a volume limited subsample drawn from the SDSS DR6.
       The multifractal analysis quantifies the scaling properties of a
       number of different moments of the galaxy counts, allowing for the
       possibility that this may differ depending on the density
       environment. The current work improves upon the earlier 2D analysis of
       \citet{2005MNRAS.364..601Y}. The large contiguous region covered by
       SDSS DR6 allows us to carry out a full three dimensional analysis
       covering scales upto $103 \,h^{-1} {\rm Mpc}$ in the volume limited
       sample.  The transition to homogeneity has been identified by comparing 
       the observations with a random point distribution having the same number
       and geometry as the actual galaxy distribution. The observations have also 
       been compared with $\Lambda$CDM N-body simulations which are also used 
       to determine the error-bars.

       A brief outline of the paper follows. We describe the data and the
       method of analysis in the section \ref{dam}. Section \ref{res}
       presents the results and conclusions of our study.

       \section{Data and Method of Analysis}
       \label{dam}
       \subsection{SDSS DR6 Data}
       Our present analysis is based on galaxy redshift data from the SDSS
       DR6 \citep{adelman}.  The SDSS DR6 includes 9583 square degrees of
       imaging and 7425 square degrees of spectroscopy with $790,860$ galaxy
       redshifts.  For the present work we have used the Main Galaxy Sample
       for which the target selection algorithm is detailed in
       \citet{strauss}. The Main Galaxy Sample comprises of galaxies brighter
       than a limiting r band Petrosian magnitude $17.77$.  The data was
       downloaded from the Catalog Archive Server (CAS) of SDSS DR6 using a
       Structured Query Language (SQL) search. We have identified a
       contiguous region in the Northern Galactic Cap which spans
       $-50^\circ<\lambda<30^\circ$ and $-6^\circ<\eta<35^\circ$ where
       $\lambda$ and $\eta$ are survey co-ordinates defined in
       \citet{stout}. A volume limited galaxy subsample was constructed in
       this region by restricting the extinction corrected Petrosian r band
       apparent magnitude to the range $14.5 \leq m_r \leq 17.77$ and
       restricting the absolute magnitude to the range $-20\geq M_r \geq
       -21$. This gives us $41,234 $ galaxies in the redshift range $0.04
       \leq z \leq 0.11$ which corresponds to comoving radial distances from
       $130 \, h^{-1} {\rm Mpc}$ to $335 \, h^{-1} {\rm Mpc}$.

       \subsection{N-body Data}
       We have simulated the dark matter distribution at $z=0$ using a
       Particle-Mesh (PM) N-body code. This was used to simulate the observed
       galaxy distribution by identifying randomly chosen dark matter
       particles as galaxies.  Here we have chosen $41,234 $ particles from a
       region which is exactly identical to the survey volume.
       
       The N-body simulations use $512^3$ particles on a $512^3$ mesh, with
       grid spacing $1 \ h^{-1} {\rm Mpc}$. The simulations use a
       $\Lambda$CDM power spectrum with cosmological parameters
       $(\Omega_{m0}, \Omega_{\Lambda 0}, h, n_s, \sigma_8)$ = $(0.274,
       0.726, 0.705, 0.96, 0.812)$ \citep{komatsu08}.  The effect of peculiar
       velocities was incorporated using the plane parallel approximation.
       We have run five independent realizations of the N-body code, and the
       simulated galaxy distribution were analysed in exactly the same way as
       the actual data.
       
       The galaxies in our simulations are expected to exactly trace the dark
       matter distribution. Galaxy formation is a complex non-linear process,
       and it is quite possible that the galaxies are a biased tracer of the
       underlying dark matter distribution. To account for this possibility,
       we have also considered a semi analytic galaxy catalogue \citep{croton}
       from Millennium Run Simulation \citep{springel}. Semi analytic models
       are simplified simulations of the formation and evolution of galaxies
       in a hierarchical clustering scenario incorporating all relevant
       physics of galaxy formation processes.  The spectra and magnitude of
       the model galaxies were computed using population synthesis models of
       \citet{brujual} and we use the catalog where the galaxy magnitudes are
       available in the SDSS u, g, r, i, z filters. The catalog contains
       about 9 million galaxies in a $(500 \, h^{-1} {\rm Mpc})^{3}$ box.
       Using the peculiar velocities we map the galaxies to redshift space
       and then identify a region having same geometry and choose the same
       number of galaxies within the specified magnitude range as the actual
       galaxy sample.
       
       \subsection{Method of analysis}
       Scale invariance of galaxy clustering \citep{2005RvMP...76.1211J},
       atleast at small scales, motivates us to use fractal analysis.  There
       are various definitions of fractal dimension e.g.  the Hausdorff
       Dimension, Capacity Dimension, Similarity Dimension, Box counting
       Dimension, etc.  These definitions represent particular cases of the
       multifractal spectrum of generalised dimension. In this paper, we use
       the generalised Minkowski-Bouligand dimension to characterise the
       galaxy distribution. This contains complete information about various
       moments of the galaxy counts, and hence, is well suited for the
       multifractal analysis.
       
       Considering  the $i^{th}$  galaxy as center, we determine $n_i(<r)$ 
       the number of other  galaxies  within a sphere of comoving radius $r$.  
       The generalised correlation integral is defined as 
       \begin{equation}
         C_q(r)=\frac{1}{MN}  \sum^{M}_{i=1}[n_i(<r)]^{q-1} 
         \label{eq:GCI}
       \end{equation}
       where $M$ is the number of centers, $N$ is the total number of
       galaxies and $q-1$ refers to a particular moment of the galaxy counts.
       For a fixed $r$, we have used all the galaxies, barring those that lie
       within a distance $r$ from the survey boundary, as centers.  In our
       analysis $N=41234$ and $M = 27706$ and $6108$ for $r=20$ and $70 \
       h^{-1} \ {\rm Mpc}$ respectively.

       The generalised Minkowski-Bouligand dimension $D_q$ follows from the
       correlation integral, and is given by
       \begin{equation}
         D_q(r) = \frac{1}{q-1} \frac{d ~ \rm \log{C_q(r)}}{d ~ \rm \log{r}~ ~ ~ ~ ~~}
         \label{eq:MBD}
       \end{equation}
       For $q > 1$, $D_q$ probes the scaling behaviour of galaxies in high
       density environments e.g. clusters and super-clusters and for $q < 1$,
       $D_q$ probes the same in under-dense environments like voids.  We have
       varied $q$ from $q=-4$ to $q=+4$ in steps of 1.  In principle, we
       could have considered even higher and lower values of $q$ but the
       finiteness of the data increases the scatter in the $D_q$ value as the
       value of $\mid q \mid$ increases \citep{2008MNRAS.390..829B}. We have
       calculated $D_q(r)$ by numerically differentiating $C_q$ using the
       Ridders' method \citep{1992nrca.book.....P} considering $C_q(r)$ at
       three consecutive $r$ values each separated by $5 \,h^{-1} {\rm
         Mpc}$.  Using a smaller $r$ interval ($1\,h^{-1} {\rm Mpc}$) , we
       find an increase in the fluctuations in $D_q(r)$ while using a larger
       interval ($10 \,h^{-1} {\rm Mpc}$) gives nearly the same results as
       ($5 \,h^{-1} {\rm Mpc}$).

       We have determined $C_q(r)$ in the range $1 \le r \le 103 \,h^{-1}
       {\rm Mpc}$ and $D_q(r)$ in the range $7 \le r \le 98 \,h^{-1} {\rm
         Mpc}$ for the data as well as the simulations. To assess the
       transition to homogeneity we have also generated ``random samples''
       which contain randomly located points. The number of points and the
       volume coverage of these samples is exactly same as that of SDSS data. 
       The point distribution in the random samples are homogeneous by construction,
       and we have generated 10 independent samples. The data, or the
       simulations, are considered to be homogeneous on length-scales where
       $D_q(r)$ is consistent with that of the random sample.

       \section{Results and Conclusions}
       \label{res}

       \begin{figure*}
         \includegraphics[width=0.95\textwidth]{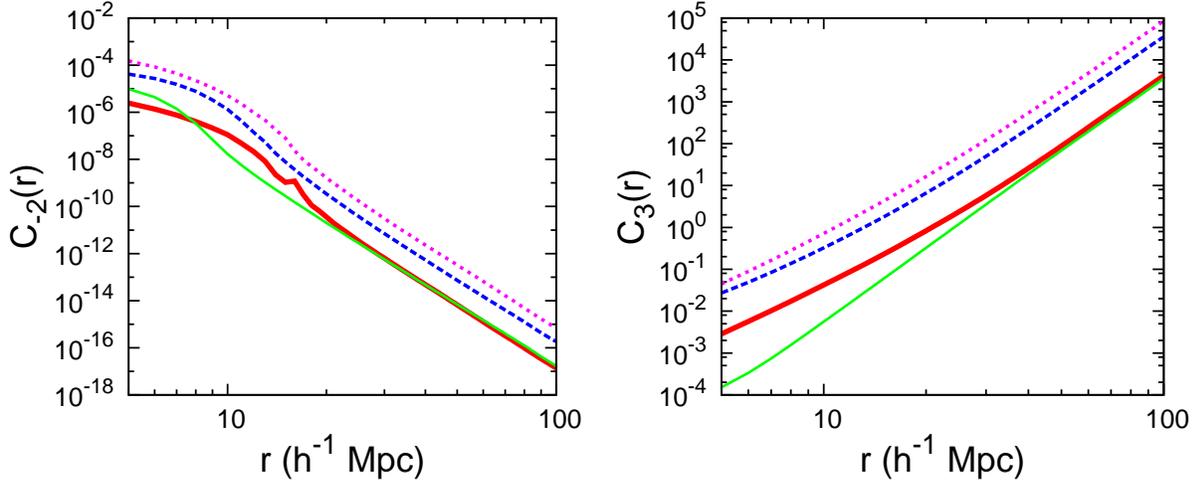}
         \caption{This shows $C_q(r)$ for $q=-2$ (left panel) and $q=3$
           (right).  The different curves (top to bottom) show the Millennium Run
           (pink-dotted), dark matter N-body (blue-dashed), SDSS data (red-thick
           solid) and random data (green-thin solid). The two simulation curves
           have been scaled arbitrarily for convenience of plotting.  The four
           curves shown here completely overlap at large $r$ if plotted with the
           same scale.  For the N-body and random data, the mean $C_q$ averaged
           over different independent realisations is shown.}
         \label{fig:cqvsr}
       \end{figure*}
    
       \begin{figure*}
         \includegraphics[width=0.95\textwidth]{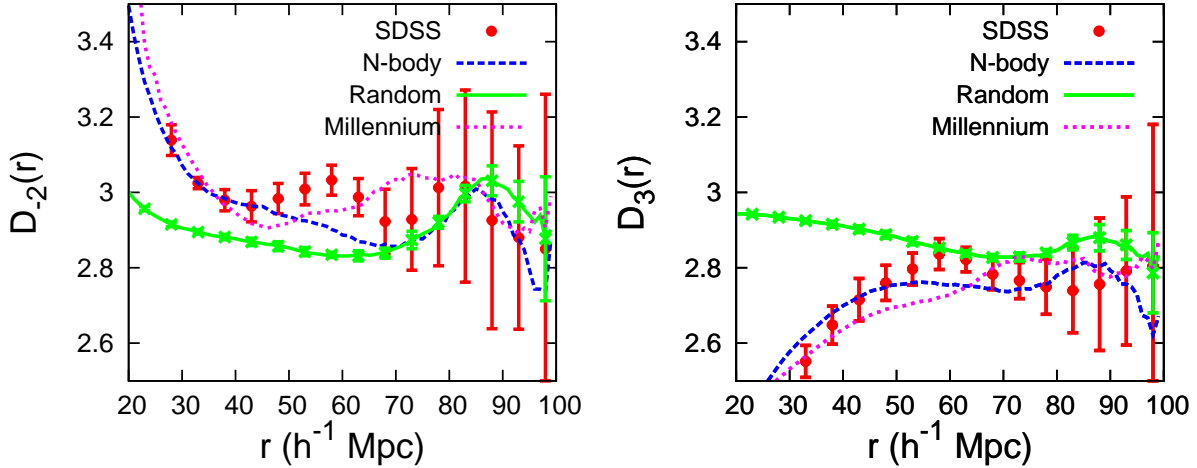}
         \caption{This shows $D_q(r)$ for $q=-2$ (left panel) and $q=3$
           (right).  The mean and $1-\sigma$ error-bars for the random data and
           the N-body simulations were respectively determined using $10$ and
           $5$ independent realisations.  We have used the N-body $1-\sigma$
           error-bars to estimate the expected statistical fluctuation in the
           actual SDSS data, and show these on the SDSS results instead of the
           mean N-body curve. We consider the SDSS data to be consistent with
           the random data if the random data lies within the $1-\sigma$
           error-bars of the SDSS.  }
         \label{fig:dqvsr}
       \end{figure*}
 
       \begin{figure*}
         \includegraphics[width=0.95\textwidth]{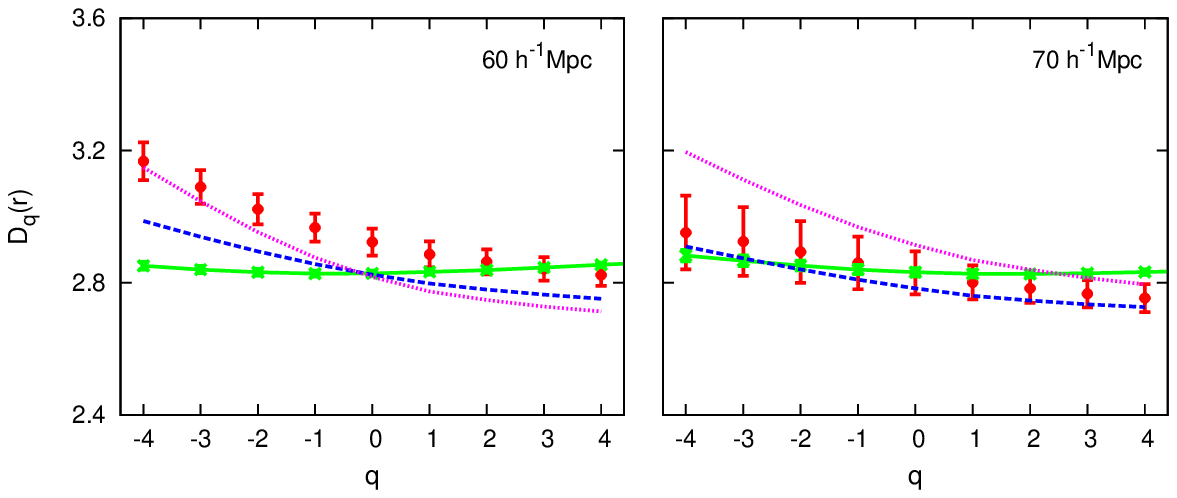}
         \caption{This shows $D_q$ as a function of $q$ for $r=60$ (left panel)
           and $70 \,h^{-1} {\rm Mpc}$ (right).  We show the SDSS data (red
           points), random data (green solid line), N-body (blue dashed) and
           Millennium (pink dotted).  Five independent N-body realisations were
           used to estimate the $1-\sigma$ error-bars shown on the SDSS data.  }
         \label{dqvsq1}
       \end{figure*}

       Figure \ref{fig:cqvsr} shows the correlation integral $C_q(r)$ for $q
       = -2$ and $q = +3$.  The behaviour is similar for other values of $q$.
       We find that $C_q(r)$ increases with $r$ for positive values of $q$
       whereas it falls progressively for negative values of $q$. It is quite
       clear that at all length-scales the SDSS galaxies, the dark matter and
       the Millennium simulations all exhibit the same scaling behaviour. It
       is also quite clear that on small scales, less than approximately $30$
       and $40 \,h^{-1} {\rm Mpc}$ for $q=-2$ and $3$ respectively, this
       scaling behaviour is distinctly different from that of the random
       samples. The scaling behaviour of the actual data, simulations and the
       random sample all appear to match as $r \rightarrow 100 \, h^{-1} \
       {\rm Mpc}$, and the $C_q(r)$ curves all coincide if plotted on the
       same scale. The match between the data and the random samples
       indicates a transition to homogeneity at a length-scale smaller than
       $100 \,h^{-1} {\rm Mpc}$, and the issue is to identify the
       length-scale where this transition occurs.

       As discussed earlier, we have numerically differentiated $C_q(r)$ to
       calculate the Minkowski-Bouligand dimension $D_q(r)$ at different
       value of $r$. The results are shown in Figure \ref{fig:dqvsr} for
       $q=-2$ and $+3$.  The behaviour is similar for other values of $q$. We
       expect the random samples to have $D_q=3$, the same as the dimension
       of the ambient space, irrespective of the value of $r$. We find that
       $D_q(r)$ has values close to $3$, the deviations being less than $7
       \%$.  At length-scales of $r \sim 30 \,h^{-1} {\rm Mpc}$ the data
       and the simulations have $D_q(r)$ values that are quite different from
       the random samples.  We find $D_q \sim 3.2$ and $2.6$ for $q=-2$ and
       $+3$ respectively. The difference from the random samples is larger
       than the $1-\sigma$ statistical fluctuations expected for the data, It
       is clear from this figure that the SDSS data is consistent with the
       random data at length-scales $70 \, h^{-1} {\rm Mpc}$ and larger. Based on
       this we conclude that the transition to homogeneity occurs at a
       length-scale between $60$ and $70 h^{-1} {\rm Mpc}$. We note that the dark
       matter N-body simulations and the Millennium simulation also exhibit a
       transition to homogeneity at a similar length-scale as the SDSS
       galaxies.

       We next consider the full spectrum of generalised dimension $D_q(r)$
       with $-4 \le q \le 4$ at $r=60$ and $70 \ h^{-1} {\rm Mpc}$ (Figure
       \ref{dqvsq1}). At $60 \,h^{-1} {\rm Mpc}$ the SDSS and random data
       are consistent for positive $q$, but they do not agree for $q \le 0$
       where $D_q$ is larger for the SDSS data.  This indicates that at $60 \
       h^{-1} \ {\rm Mpc}$ the overdense regions are consistent with
       homogeneity whereas the underdense regions are not. At $70 \,h^{-1}
       {\rm Mpc}$ we find that the SDSS and random data are consistent for
       the entire $q$ range that we have analysed. This is in keeping with
       our conclusion that the transition to homogeneity occurs at a
       length-scale between $60$ and $70 \,h^{-1} {\rm Mpc}$.
       
       Our results are in disagreement with similar analyses by
       \citet{2007A&A...465...23S,2009EL.....8529002S,2009EL.....8649001S}
       who report long range correlations and persistent fluctuations in the
       large scale galaxy distribution, and fail to find a transition to
       homogeneity.  The results of this paper are in good agreement with an
       earlier 2D analysis of the SDSS main galaxy sample
       \citep{2005MNRAS.364..601Y} and the SDSS LRG sample
       \citep{2005ApJ...624...54H} both of which find a transition to
       homogeneity at around $70 \,h^{-1} {\rm Mpc}$. Note that the LRG
       sample covers a much larger volume as compared to the main galaxy
       sample considered in this paper. A visual inspection of the galaxy
       distribution reveals that the galaxies appear to be distributed in an
       interconnected network of filaments encircling voids. This network,
       referred to as the ``Cosmic Web'', appears to fill the entire region
       covered by galaxy surveys. \citet{bh1} and \citet{pandey} have shown
       that the observed filaments are statistically significant only to
       length-scales of $70 \, h^{-1} \ {\rm Mpc}$ and not beyond.  Longer
       filaments, though seen, are produced by chance alignments and are not
       statistically significant.  The good agreement between all of these
       findings provide strong evidence that the galaxy distribution exhibits
       a transition to homogeneity at around $70 \,h^{-1} {\rm Mpc}$.

       \section*{Acknowledgment}
       BP acknowledges Darren Croton for his help in analysing the
       Millennium catalogue. P.S. acknowledges financial support of 
       University Grant Commission, India. JY thanks J. S. Bagla,
       T. R. Seshadri, S. Borgani \& F. S. Labini for useful discussions.
       
       The SDSS DR6 data was downloaded from the SDSS skyserver
       http://cas.sdss.org/dr6/en/.        
       Funding for the creation and distribution of the SDSS Archive has
       been provided by the Alfred P. Sloan Foundation, the Participating
       Institutions, the National Aeronautics and Space Administration,
       the National Science Foundation, the U.S. Department of Energy,
       the Japanese Monbukagakusho, and the Max Planck Society. The SDSS
       Web site is http://www.sdss.org/.
       
       The SDSS is managed by the Astrophysical Research Consortium (ARC)
       for the Participating Institutions. The Participating Institutions
       are The University of Chicago, Fermilab, the Institute for
       Advanced Study, the Japan Participation Group, The Johns Hopkins
       University, the Korean Scientist Group, Los Alamos National
       Laboratory, the Max-Planck-Institute for Astronomy (MPIA), the
       Max-Planck-Institute for Astrophysics (MPA), New Mexico State
       University, University of Pittsburgh, Princeton University, the
       United States Naval Observatory, and the University of Washington.
       
       The Millennium Run simulation used in this
       paper was carried out by the Virgo Supercomputing Consortium at
       the Computing Centre of the Max-Planck Society in Garching. The
       semi-analytic galaxy catalogue is publicly available at
       http://www.mpa-garching.mpg.de/galform/agnpaper.

\end{document}